\pdfoutput=1	% for arXiv

\documentclass[10pt,letterpaper,twocolumn]{article} %% two column, final layout

\usepackage{ol2}
\usepackage[draft]{hyperref}
\usepackage{amsmath}

\begin{document}

% HERE...

% ... EITHER uncomment the next two lines to see the changes highlighted...
%\newcommand{\change}[2]{\textbf{#1 [was ``#2'']}}
%\newcommand{\add}[1]{\textbf{#1}}
%\newcommand{\cut}[1]{\textbf{[deleted ``#1'']}}

% ... OR uncomment the next two lines to make the changes appear like normal text
\newcommand{\change}[2]{#1}
\newcommand{\add}[1]{#1}
\newcommand{\cut}[1]{}

\renewcommand{\section}[1]{}

\twocolumn[ %% activate for two-column option

\title{\change{Fermat's principle and the formal equivalence of local light-ray rotation and refraction at the interface between homogeneous media with a complex refractive-index ratio}{Fermat's principle with complex refractive indices and local light-ray rotation}}

\author{Bhuvanesh Sundar, Alasdair C.\ Hamilton,$^1$ and Johannes Courtial}
\address{
Department of Physics and Astronomy, Faculty of Physical Sciences, University of Glasgow, Glasgow, UK \\
$^1$Corresponding author: a.c.hamilton@physics.gla.ac.uk
}

\begin{abstract}
We \change{derive a formal description of}{describe} local light-ray rotation in terms of complex refractive indices.
We show that Fermat's principle holds, and we derive an extended Snell's law.
The change in the angle of a light ray with respect to the normal to a refractive-index interface is described by the modulus of the refractive-index ratio, the rotation around the interface normal is described by the argument of the refractive-index ratio.
\end{abstract}

%% see http://www.aip.org/pacs/
%\pacs{
%% 01.50.Wg, % (Physics of toys)
%42.15.-i, % (Geometrical optics)
%42.15.Dp, % (Wave fronts and ray tracing)
%42.25.-p, % (Wave optics)
%42.25.Gy, % (Edge and boundary effects; reflection and refraction)
%42.70.-a% (Optical materials)
%}


] %% activate for two-column option

\newcommand{\rmi}{\mathrm{i}}
\newcommand{\rmd}{\mathrm{d}}
\newcommand{\bi}[1]{\mathbf{#1}}

\section{Introduction}

We recently started to investigate ray-optical analogs of metamaterials.
Like \cut{real} metamaterials \cite{Smith-et-al-2004,Pendry-et-al-2006}, \change{these so-called METATOYs (\underline{meta}ma\underline{t}erials f\underline{o}r ra\underline{y}s) \cite{Hamilton-Courtial-2009}}{was ``our'' metamaterials for rays (``METATOYs'')} are capable of performing positive and negative refraction \cite{Courtial-Nelson-2008,Courtial-2008a}.
\change{In addition}{Unlike \cut{real} metamaterials}, METATOYs can \cut{also} perform local light-ray rotation around the interface normal \cite{Hamilton-et-al-2009}\cut{, a concept without wave-optical analog \cite{Hamilton-Courtial-2009}}.

Here we \cut{attempt to} describe local light-ray rotation \add{around the interface normal} in terms of Fermat's principle.
% \cut{As Fermat's principle is very much based on wave optics, this seems futile.}
\change{Fermat's principle can be treated as a basic theorem of geometrical optics, but ``is itself only understandable in terms of a wave theory'' \cite{Smith-Thomson-1988-Fermat}.
On the other hand, local light-ray rotation around the interface normal has no wave-optical analog in the sense that it is not always possible to construct a wave in which the phase-front normal -- the geometrical-optics light-ray direction -- has been rotated as required \cite{Hamilton-Courtial-2009}.
What, then, happens if we attempt to describe local light-ray rotation around the interface normal in terms of Fermat's principle?}{As Fermat's principle is very much based on wave optics, this seems futile.}
\add{Here we do just that.}
\change{We}{Nevertheless, we} find a ``natural'' formulation of Fermat's principle in which ray rotation is described by an interface between \add{homogeneous} media with a complex refractive-index ratio.
This leads to an extended form of Snell's law that uses complex refractive indices.

It is important to note that the \add{meaning of the imaginary part of the} complex refractive index we introduce here is \cut{very} different from that commonly used in optics\change{ \cite{Born-Wolf-1980-complex-refractive-index}:
in the former it is associated with ray rotation, in the latter with attenuation.}{, where the imaginary part of the complex refractive index describes attenuation \cut{\cite{Born-Wolf-1980-complex-refractive-index}}.
In contrast, here the argument of the complex refractive index is related to the angle by which light rays are rotated around the interface normal as they pass from one medium to another.}

% This paper is organized as follows.

\section{Fermat's principle at a planar interface}
Fermat's principle \cite{Born-Wolf-1980-Fermat} states that a light ray traveling between two points takes a path with a stationary optical path length, that is, for small variations in the path taken the optical path length stays the same.
The whole path can be calculated using the calculus of variations, but simplified calculations can be performed using ordinary calculus.

\begin{figure}[hb]
\begin{center}
\includegraphics{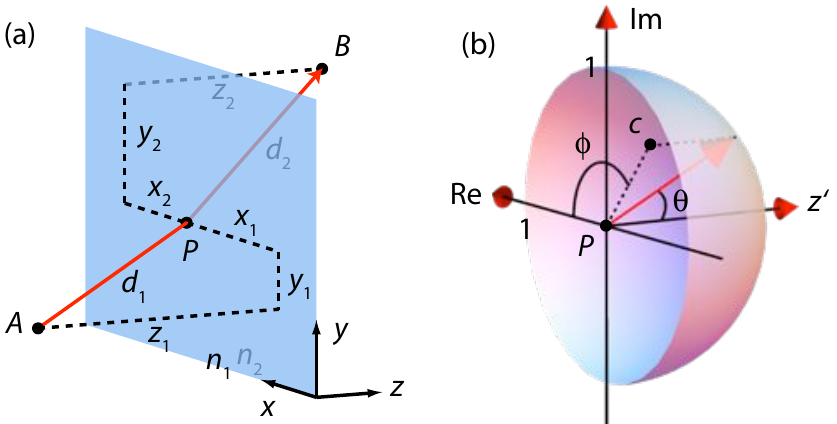}
\end{center}
\caption{\label{geometry-figure}(Color online)
Geometry of refraction at a planar interface between two media with different refractive indices, $n_1$ and $n_2$.
(a)~A light ray travels from a point $A$ in front of the interface to a point $P$ on the interface, and then to a point $B$ behind the interface.
The geometrical distance between $A$ and $P$ is $d_1$, that between $P$ and $B$ is $d_2$.
(b)~The light-ray direction can be represented by two angles, $\theta$ and $\phi$, which respectively represent the angle with respect to the interface normal (the $z$ axis) and the angle of the projection into the interface plane with respect to the $x$ axis.
Alternatively, the light-ray direction can be described by the projection of the normalized direction vector into the interface plane.
With \change{a complex}{an Argand} plane in the interface plane as shown, this projection can then be described by a single complex number, $c$.}
\end{figure}

Figure \ref{geometry-figure}(a) shows a planar interface between two media with different refractive indices, $n_1$ and $n_2$.
When light travels between two fixed points $A$ and $B$ on either side of the interface, via a point $P$ on the interface, but not otherwise fixed, it covers an optical path length
\begin{eqnarray}
\Delta &=& n_1 d_1 + n_2 d_2 \\
&=& n_1 \sqrt{x_1^2 + y_1^2 + z_1^2} + n_2 \sqrt{x_2^2 + y_2^2 + z_2^2}.
\label{path-difference-equation}
\end{eqnarray}
$(x_1, y_1, z_1)$ and $(x_2, y_2, z_2)$ are the components of the vector $AP$ (that is, the vector from $A$ to $P$) and $PB$, respectively, in a Cartesian coordinate system whose $(x,y)$ plane coincides with the plane of the interface.
We can use the equations
\begin{equation}
X = x_1 + x_2, \quad Y = y_1 + y_2,
\end{equation}
which describe the constant separations between the fixed points $A$ and $B$ in the $x$ and $y$ directions, to eliminate $x_2$ and $y_2$ from the expression for the path difference, Eqn (\ref{path-difference-equation}).
This gives
\begin{eqnarray}
\Delta &=& n_1 \sqrt{x_1^2 + y_1^2 + z_1^2} + \nonumber \\
&& n_2 \sqrt{(X-x_1)^2 + (Y-y_1)^2 + z_2^2}.
\label{path-difference-equation-1}
\end{eqnarray}

We now find the values $x_1$ and $y_1$ for which the function $\Delta$ is stationary.
According to Fermat's principle, these values then correspond to the point $P$ through which the light ray would actually travel.
These values $x_1$ and $y_1$ have to satisfy the equations
\begin{equation}
\frac{\partial \Delta}{\partial x_1} = 0, \quad
\frac{\partial \Delta}{\partial y_1} = 0.
\end{equation}
Substitution of the expression for $\Delta$ into these equations gives
\begin{equation}
n_1 \frac{x_1}{r_1} - n_2 \frac{x_2}{r_2} = 0, \quad
n_1 \frac{y_1}{r_1} - n_2 \frac{y_2}{r_2} = 0,
\label{x-and-y-derivative-conditions}
\end{equation}
where $r_j = \left( x_j^2 + y_j^2 + z_j^2 \right)^{1/2}$ (with $j=1, 2$).
The terms can be translated into spherical coordinates $\phi$ (the azimuthal angle) and $\theta$ (the angle with the $z$ axis -- see Fig.\ \ref{geometry-figure}(b)) using the equations
\begin{equation}
\frac{x_j}{r_j} = \sin \theta_j \cos \phi_j, \quad
\frac{y_j}{r_j} = \sin \theta_j \sin \phi_j.
\end{equation}
In spherical coordinates, equations (\ref{x-and-y-derivative-conditions}) are therefore the real and imaginary part, respectively, of the equation
\begin{equation}
n_1 \sin \theta_1 \exp (\rmi \phi_1) = n_2 \sin \theta_2 \exp(\rmi \phi_2).
\label{extended-Snell-equation}
\end{equation}
% (We were using $\cos \phi_j + \rmi \sin \phi_j = \exp( \rmi \phi_j)$.)

This equation is the basis of the remainder of this paper.
Like in Snell's law, it is not the individual refractive indices that matter, but their ratio.
As equation (\ref{extended-Snell-equation}) is a complex equation, it is natural to allow the refractive indices (and their ratio) to be complex numbers.
Equation (\ref{extended-Snell-equation}) is then an extension of Snell's law:
for real refractive-index ratios, it describes ordinary refraction according to Snell's law, expressing (unlike the Snell's-law formula) the fact that the refracted ray lies in the same plane as the incident ray and the interface normal;
and it leads to local light-ray rotation as a natural extension of refraction with complex refractive-index ratios.
We discuss these properties below.

\section{Geometric interpretation}
For the following discussion it is useful to visualize the extended Snell's law as follows.
We interpret the plane of the refractive-index interface as a complex plane, centered at the point $P$ where the light ray intersects the plane, and with the real axis in the $x$ direction and the imaginary axis in the $y$ direction.
$z^\prime$ is the normal to the interface at $P$.
Figure \ref{geometry-figure}(b) shows this coordinate system.
% The $z$ axis is then the interface normal at the intersection point.

We now consider a unit vector in the direction of the incident light ray, starting at the origin.
We define the complex number $c$ as the orthographic projection of this unit vector into the complex plane.
If we assume that light travels in the positive $z'$ direction, then this projection uniquely defines the ray direction.
For a unit vector with spherical-coordinate angles $\theta$ and $\phi$, $c$ is
\begin{equation}
c = \sin \theta \exp (\rmi \phi).
\end{equation}

The extended Snell's law, equation (\ref{extended-Snell-equation}), can then be written in the form
\begin{equation}
n_1 c_1 = n_2 c_2,
\label{n-c-equation}
\end{equation}
where $c_1$ and $c_2$ is the complex number corresponding to the incident and refracted ray direction, respectively.
In other words, the product of refractive index and the complex number representing the ray direction remains constant.
We can emphasize the dependence on the refractive-index ratio by writing equation (\ref{n-c-equation}) in the form
\begin{equation}
\frac{n_1}{n_2} = \frac{c_2}{c_1}.
\label{ratio-equation}
\end{equation}

% We discuss examples below.

\begin{figure}
\begin{center}
\includegraphics{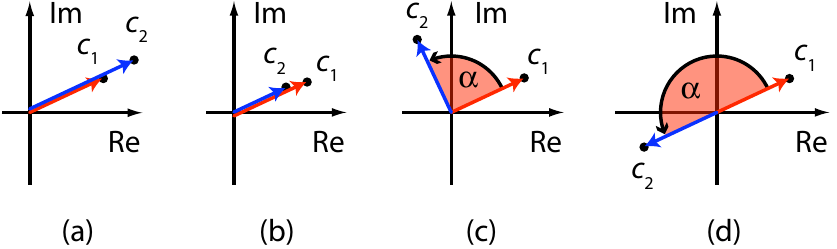}
\end{center}
\caption{\label{c-plots-figure}(Color online)
Plots of the complex numbers $c_1$ and $c_2$ representing various types of refraction.
(a) and (b) are examples of standard refraction ($n_1/n_2$ real and positive).
In (a) $n_1/n_2 > 1$, $n_1/n_2 < 1$ in (b).
(c)~A complex refractive-index ratio $n_1/n_2 = \exp(\rmi \alpha)$ leads to local light-ray rotation through an angle $\alpha$ (here $\alpha = 90^\circ$).
(d)~Light-ray rotation through $180^\circ$ is equivalent to negative refraction with a refractive-index ratio $n_1/n_2 = -1$.}
\end{figure}

\section{Interface between media with real refractive-index ratio}
According to equation (\ref{ratio-equation}), the ratio $c_2/c_1$ is the same as the refractive-index ratio $n_1 / n_2$.
Therefore, if the refractive-index ratio is real then so is the ratio of the direction projections.

This has a simple interpretation.
The argument of $c_1$ -- the spherical-coordinates angle $\phi_1$ -- defines the plane of incidence.
Specifically, it describes the angle between the plane of incidence and the $x$ axis.
This, together with the fact that the plane of incidence also contains the $z$ axis, completely determines the plane of incidence.
A real ratio $c_2/c_1$ means that $c_2$ lies on the same line through the origin as $c_1$ (Fig.\ \ref{c-plots-figure}(a) and (b)), and so the refracted ray is also in the plane of incidence.

Mathematically, it means that $\phi_1 = \phi_2$.
This means that the $\exp(\rmi \phi_j)$ terms ($j=1,2$) can be cancelled in equation (\ref{extended-Snell-equation}), which reduces to Snell's law,
\begin{equation}
n_1 \sin \theta_1 = n_2 \sin \theta_2.
\label{Snell-equation}
\end{equation} 
% So the extended Snell's law, equation (\ref{extended-Snell-equation}), expresses both the fact 

\section{Interface between media with complex refractive-index ratio}
\change{Perhaps the}{The} simplest examples of complex refractive-index ratios are those of the form
\begin{equation}
\frac{n_1}{n_2} = \exp(\rmi \alpha)
\end{equation}
(where $\alpha$ is a real number).
According to equation (\ref{ratio-equation}), this implies that
\begin{equation}
\frac{c_2}{c_1} = \exp(\rmi \alpha).
\end{equation}
This means that $c_2$, which characterizes the projection of the refracted ray into the interface plane, is rotated with respect to $c_1$ through an angle $\alpha$ around the point $P$ (Fig.\ \ref{c-plots-figure}(c)).
In three dimensions, it means that the direction of the refracted ray is that of the incident ray, rotated around the $z^\prime$ axis through an angle $\alpha$.
This is precisely the local light-ray rotation that can be achieved with METATOYs \cite{Hamilton-et-al-2009,Hamilton-Courtial-2009}.

The case of general complex refractive-index ratios $n_1/n_2$ can be approached by writing the left-hand side of equation (\ref{ratio-equation}) in terms of the modulus and argument of this ratio, namely
\begin{equation}
\frac{n_1}{n_2} = \left| \frac{n_1}{n_2} \right| \exp \! \left[ \rmi \arg \! \left( \frac{n_1}{n_2} \right) \right],
\label{lhs-equation}
\end{equation}
and the right-hand side in the form
\begin{equation}
\frac{\sin \theta_2 \exp(\rmi \phi_2)}{\sin \theta_1 \exp(\rmi \phi_1)} =
\frac{\sin \theta_2 \exp[\rmi (\phi_1 + \alpha)]}{\sin \theta_1 \exp(\rmi \phi_1)} =
\frac{\sin \theta_2}{\sin \theta_1} \exp(\rmi \alpha),
\label{rhs-equation}
\end{equation}
which \change{expresses}{assumes that} the direction of the refracted light ray's projection \change{as}{is} that of the incident light ray, rotated through an angle $\alpha$ around $P$.
Comparison of the moduli of equations (\ref{lhs-equation}) and (\ref{rhs-equation}) reveals that the change of the angle between the ray and the $z^\prime$ axis is then given by the absolute value of the refractive-index ratio according to the equation
\begin{equation}
\left| \frac{n_1}{n_2} \right| = \frac{\sin \theta_2}{\sin \theta_1},
\label{absolute-part-equation}
\end{equation}
which, for real and positive refractive-index ratios, is the same as Snell's law (Eqn.\ (\ref{Snell-equation})).
Comparison of the arguments reveals that the rotation angle $\alpha$ is given by the argument of the refractive-index ratio:
\begin{equation}
\arg \! \left( \frac{n_1}{n_2} \right) = \alpha.
\label{argument-part-equation}
\end{equation}

\section{Negative refraction}

Now we discuss briefly the case of negative refraction \cite{Pendry-Smith-2004}, for which the refractive-index ratio $n_1/n_2$ is real and negative.
Negative refraction is fully described by Snell's law (and indeed \change{the}{our} extended Snell's law), from which it then follows that the angle of the ray with the $z^\prime$ axis, $\theta$, has to change sign.
It can alternatively be described by a change of the angle $\theta$ without a sign change in combination with a ray rotation around the $z^\prime$ axis through $180^\circ$.
Mathematically, the equivalence between negative refraction and ray rotation through $180^\circ$ can be expressed as
\begin{equation}
c = \sin (-\theta) \exp(\rmi \phi) = \sin \theta \exp(\rmi (\phi + 180^\circ)).
\end{equation}

The case $n_1/n_2 = -1$ is shown in Fig.\ \ref{c-plots-figure}(d).
In fact, a Dove-prism-array structure that is ray-optically equivalent to a refractive-index interface with $n_1/n_2 = -1$ \cite{Courtial-Nelson-2008} is a special case of a Dove-prism-array ray-rotator \cite{Hamilton-et-al-2009} for rotation angle $\alpha = 180^\circ$.

\section{Conclusions}
Ray rotation is a concept that has no wave-optical analog.
It is curious that it is possible to describe it -- in such a natural manner -- by using \cut{the wave-optical} Fermat's principle, albeit with a complex refractive-index ratio.
More work may lead to a deeper understanding.

%\begin{itemize}
%\item Alternatively, can the complex exponential somehow be absorbed in the sine term by making the angles $\theta_j$ complex?
%\end{itemize}

\section*{Acknowledgments}

ACH is supported by the UK's Engineering and Physical Sciences Research Council (EPSRC).
JC is a Royal Society University Research Fellow.

% \section*{References}

\bibliographystyle{osajnl}
\bibliography{/Users/johannes/Documents/work/library/Johannes}

\end{document}